\documentclass[sigconf]{acmart}
\AtBeginDocument{%
  }

\usepackage{enumitem}
\usepackage{calc} 
\usepackage{balance}
\usepackage{graphicx}
\usepackage{subcaption}

\copyrightyear{2025}
\acmYear{2025}
\setcopyright{acmlicensed}\acmConference[SIGIR '25]{Proceedings of the 48th International ACM SIGIR Conference on Research and Development in Information Retrieval}{July 13--18, 2025}{Padua, Italy}
\acmBooktitle{Proceedings of the 48th International ACM SIGIR Conference on Research and Development in Information Retrieval (SIGIR '25), July 13--18, 2025, Padua, Italy}
\acmDOI{10.1145/3726302.3730305}
\acmISBN{979-8-4007-1592-1/2025/07}

\begin{document}

\title{Benchmarking LLM-based Relevance Judgment Methods}


\author{Negar Arabzadeh}
\email{narabzad@uwaterloo.ca}
\orcid{0000-0002-4411-7089}
\affiliation{%
  \institution{University of Waterloo}
  \city{Waterloo}
  \state{Ontario}
  \country{Canada}
}

\author{Charles L.A. Clarke}
\email{claclark@uwaterloo.ca}
\orcid{0000-0001-8178-9194}
\affiliation{%
  \institution{University of Waterloo}
  \city{Waterloo}
  \state{Ontario}
  \country{Canada}
}

\renewcommand{\shortauthors}{Trovato et al.}

\begin{abstract}
Large Language Models (LLMs) are increasingly deployed in both academic and industry settings to automate the evaluation of information seeking systems, particularly by generating graded relevance judgments. Several studies report Kendall $\tau$ correlations exceeding $0.85$ when comparing system rankings derived from human versus LLM-generated relevance labels. Previous work on LLM-based relevance assessment has primarily focused on replicating graded human relevance judgments through various prompting strategies. However, there has been limited exploration of alternative assessment methods or comprehensive comparative studies. In this paper, we systematically compare multiple LLM-based relevance assessment methods, including binary relevance judgments, graded relevance assessments, pairwise preference-based methods, and two nugget-based evaluation methods~--~document-agnostic and document-dependent. Wherever possible, we employ state-of-the-art tools and optimized prompts tailored for these methods. In addition to a traditional comparison based on system rankings using Kendall correlations, we also examine how well LLM judgments align with human preferences, as inferred from relevance grades. We conduct extensive experiments on datasets from three TREC Deep Learning tracks 2019, 2020 and 2021 as well as the ANTIQUE dataset, which focuses on non-factoid open-domain question answering. Beyond dataset-specific results, our work offers a practical methodology for evaluating diverse LLM-based relevance assessment methods. As part of our data release, we include relevance judgments generated by both an open-source (Llama3.2b) and a commercial (gpt-4o)  model.  Our goal is to \textit{reproduce} various LLM-based relevance judgment methods to provide a comprehensive comparison. We release all the relevance judgments as a \textit{resource} that establishes a baseline for future work, ensuring a level playing field for evaluation of LLM-based relevance judgments. All code, data, and resources are publicly available in our GitHub Repository at 
\url{https://github.com/Narabzad/llm-relevance-judgement-comparison}

\end{abstract}


\maketitle

\section{Introduction}

Large Language Models (LLMs) have emerged as transformative tools in the information retrieval (IR) domain \cite{zhu2023large}, enabling significant advancements in tasks such as document ranking \cite{zhang2023rank,sun2023chatgpt,qin2023large}, query generation \cite{wang2023query2doc,staudinger2024reproducibility}, and system evaluation \cite{alaofi2024generative,rahmani2024llm4eval,salemi2024evaluating}. Their ability to process and generate human-like text has enhanced efficiency and scalability across tasks that were traditionally human-led. Notably, LLMs have shown substantial promise in automating the evaluation of IR systems by generating relevance judgments that closely align with human assessments \cite{thomas2023large,li2024generation,zhuang2023beyond}.

Under the traditional IR paradigm, search engines respond to user queries with ranked lists of documents from a pre-defined corpus. The effectiveness of a search engine’s ranking is evaluated based on its ability to place relevant documents higher on a list \cite{10.1111/j.1083-6101.2007.00351.x,page1999pagerank}. Historically, this evaluation methodology relied on human annotators to label query-document pairs for relevance, quality, and other attributes. Unfortunately, this approach is resource-intensive and can suffer from issues of incomplete labeling, inconsistencies among annotators, and high costs, particularly in academic setting \cite{saracevic2008effects,upadhyay2024llmspatchmissingrelevance,zhu2022relevance}. Experimental initiatives such as TREC, NTCIR, CLEF, and FIRE have aimed to address these issues by creating reusable test collections with predefined relevance labels, allowing consistent comparison of retrieval systems~\cite{10.1007/978-3-642-04447-2_2,kando1999overview,harman1993first}.

Recent advancements in LLMs have shifted this paradigm by offering a scalable and cost-effective alternative for generating relevance labels \cite{upadhyay2024largescalestudyrelevanceassessments,pradeep2024initial,arabzadeh2024comparison,meng2024queryperformancepredictionusing}.
Studies have shown that LLM-based relevance assessments can achieve accuracy comparable to, or even exceeding, that of human annotators, as demonstrated in real-world systems like Bing \cite{thomas2023large}.
These models not only reduce labeling costs but also address "assessment holes" where unjudged results in traditional test collections lead to performance underestimation of newer models \cite{arabzadeh2022shallow}.
However, the growing reliance on LLMs for evaluation introduces challenges of consistency, alignment, and transparency \cite{bigdeli2024evaluating,arabzadeh2024offline,azzopardi2024report,arabzadeh2023adele,arabzadeh-clarke-2024-frechet,arabzadeh2023quantifying}.
As \citet{faggioli2023perspectives} note, evaluations in IR must remain grounded in human judgment to maintain trust and reliability.

In fact, despite the success of LLM relevance assessments, several papers have raised concerns about their robustness and their alignment with human preferences.
\citet{fooled} report a ``tendency for many LLMs to label passages that include the original query terms as relevant.''
They demonstrate that merely injecting query terms into a document or even adding an explicit statement that the document is relevant can influence an LLM to label the document as relevant.
\citet{clarke2024llmbasedrelevanceassessmentcant} report an experiment demonstrating that LLM assessment can be biased toward LLM-based re-ranking methods.
The authors of the \texttt{UMBRELA} assessment tool~\cite{upadhyay2024umbrela,upadhyay2024largescalestudyrelevanceassessments}--which was used for automated evaluation of TREC RAG 2024-- found that when asking LLMs and humans to rate relevance using graded values, humans tend to apply stricter criteria.
For instance, when analyzing three years of TREC Deep Learning track data (2019–2021) using \texttt{UMBRELA} and comparing it to human judgments, it was observed that on average humans labeled over 13\% more documents as non-relevant (14,961 vs.\ 17,376). Conversely, LLMs were more lenient in judging relevance, labeling over 26\% more documents as perfectly relevant (3,063 vs.\ 2,429).
This observation suggests that LLMs may interpret relevance more liberally than humans.
Differences between human and LLM-based relevance judgments are also reflected in statistical measures of alignment with human labels and agreement with system rankings \cite{10.1145/290941.291009,upadhyay2024umbrela,LLMJudge,faggioli2023perspectives}. To the best of our knowledge, the extent to which these discrepancies impact the interpretation of evaluation results across different LLM-based evaluation methods has not been extensively studied. This gap may be due to the lack of available data and the limited diversity of LLM-based relevance judgments for a given dataset.

Given the flexibility of LLMs in language processing tasks, we are no longer constrained to standard methods for relevance assessment.
Methods that once required substantial human labor~---~which may have impeded their adoption~--~now become feasible due to the reduced costs associated with LLM-based assessment.
However, there has been limited exploration of alternative methods or a comprehensive comparative studies of methods beyond graded relevance assessment.

With the goal of placing different relevance judgements on a ``level playing field'' this paper compares various 
LLM-based relevance assessment methods, including: \begin{enumerate}
    \item traditional relevance judgments in both \texttt{Binary}~\cite{faggioli2023perspectives} and graded forms (\texttt{UMBRELA})~\cite{upadhyay2024umbrela}
    \item two nugget-based methods~--~one document-agnostic (\texttt{Exam}) \cite{farzi2024exambased, sander2021exam} and  one document-dependent (\texttt{AutoNuggetizer})~\cite{pradeep2024initial}
    \item  a pairwise preference-based relevance judgment method.
\end{enumerate}

\noindent To minimize the influence of prompt engineering and enhance reproducibility, as much as possible we rely on established tools and prompts.
Our experiments are conducted using both a commercial LLM (\textit{ChatGPT-4o}) and an open-source LLM (\textit{Llama 3.2B}).
Due to space limitations, we do not present or analyze the llama results in this paper, but we include all of the data, results and analysis in our GitHub repository.  

Our of our objectives in this paper is to help comparing different LLM-based relevance judgment systems on a level playing field by considering two factors:
\begin{enumerate}[leftmargin=*] 
\item \textbf{Alignment with human labels}: 
The degree to which LLM-generated relevance judgments reflect the document ordering imputed by human labels.
\item \textbf{Agreement with system rankings}: 
The consistency between system rankings produced using LLM-generated labels and those derived from human labels.
\end{enumerate}
As observed earlier, LLMs can be more lenient in judging relevance.
The first factor (``alignment'') allows us to side-step this issue by focusing on the order in which documents are placed by the various assessment methods.
For example, if human assessment assigns a higher grade to one document vs.\ another, we expect LLM-based assessment to provide a consistent ordering, even if the LLM assigns different grades.
Although the various assessment methods express relevance in different ways, e.g., grades vs.\ nuggets, alignment with human labels allows us to make a direct comparison between different relevance judgment methods.
Full details are provided in Section~\ref{sec:HA-explain}.
For similar reasons, for the second factor we employ an flexible evaluation measure ``compatibility'' that naturally adapts to any relevance assessment method~\cite{10.1145/3340531.3411915}. Compatibility essentially compares a given ranking to the ideal permutation of judgments. Since it operates on relative relevance rather than absolute values, it enables meaningful comparisons between different relevance judgment approaches. Further details on how compatibility is computed and how we apply it in our study are provided in Section \ref{sec:SR-explain}.

We report experiments evaluating the various assessment methods in terms of these two factors.
Our study employs three years of TREC test collections (2019, 2020, and 2021) \cite{trecdl2019,trecdl2020,trecdl2021}, which are based on two different corpora: MS MARCO v1 and v2 \cite{nguyen2016ms}.
We choose these test collections since they have been employed in previous comparisons of LLM-based assessments~\cite{faggioli2023perspectives,upadhyay2024umbrela} allowing our results to be compared with previous results.
In addition, as a test of generalizability, we report experiments on the ANTIQUE dataset \cite{hashemi2020antique}, an open-domain question-answering dataset derived from Yahoo Answers.
All data and code are publicly available in our GitHub repository.
As new large language models appear and assessment methods are proposed, these judgments provide a baseline for evaluation.

This paper makes several key contributions to the study of LLM-based relevance assessment:

\begin{enumerate}[leftmargin=*]
\item We curated a comprehensive resource of relevance judgments generated using five distinct methods—binary relevance, graded relevance, pairwise preference, document-agnostic nugget-based evaluation, and document-dependent nugget-based evaluation across two large language models (GPT-4o and Llama 3.2B) and four benchmark datasets (TREC DL 2019, TREC DL 2020, TREC DL 2021, and ANTIQUE). All relevance judgments, along with code and experimental results, are publicly available on our GitHub repository\footnote{\url{https://github.com/Narabzad/llm-relevance-judgement-comparison}}.
    
    \item We systematically compare LLM-generated relevance judgments with human labels, ensuring a fair and consistent evaluation by placing them on the same playing field.
    
    \item We introduce a method for assessing how well different relevance judgment approaches align with system rankings, providing a standardized framework for cross-method comparison.

\end{enumerate}

\section{Background}
The continuing rapid advancement of AI technologies necessitates the ongoing development of evaluation benchmarks~\cite{zheng2023judging,guo2023evaluating}. Traditional benchmarks often become inadequate as models achieve high performance on them~\cite{mcintosh2024inadequacieslargelanguagemodel}. Additionally, the increasing speed of benchmarking and the emergence of novel tasks make relying solely on human annotation increasingly challenging~\cite{chang2024survey}. Automated evaluation methods have thus gained popularity, accelerating the evaluation process and facilitating tool development \cite{chern2024can,arabzadeh-etal-2024-assessing}.

While automated evaluation metrics offer scalability, human judgment remains essential for capturing subjective quality measures and ensuring alignment with human labels~\cite{wang2023aligning,amirizaniani2024developing}.
To enable effective hybrid evaluations, automated methods, particularly those based on LLMs, must be aligned with human judgment \cite{jones1995evaluating}. Furthermore, conclusions drawn from annotations should be interpretable for system comparison. For instance, in LLM-based relevance judgments for information retrieval evaluation, it is crucial to understand how these judgments influence system rankings and measure system effectiveness.
LLM-based relevance assessment is not only valuable for assessing IR systems but may also play a crucial role in evaluating retrieval-augmented generation (RAG) systems, where answers are not necessarily derived from a fixed collection \cite{zhang2024large}.

There have long been attempts to automate the process of relevance judgment 
\cite{makary2017using,makary2016towards,ravana2015ranking,soboroff2001ranking}.
However, with the emergence of LLMs, automated relevance judgment has gained significant attention, with examples like the use of automated evaluation methods in TREC RAG 2024 track~\cite{upadhyay2024largescalestudyrelevanceassessments,pradeep2024initial}. 
LLM-based relevance judgment was pioneered by \citet{faggioli2023perspectives}, where the authors discussed the potential for various levels of collaboration between humans and machines in evaluating IR systems.
\citet{thomas2023large} illustrate that LLM-based relevance judgments have on-par quality as human annotators and they could even be deployed at an industrial scale, as implemented at Microsoft Bing. 
While these studies reported a high rank-based Kendall's $\tau$ correlation (typically above 0.85) between human and LLM-graded judgments~\cite{abbasiantaeb2024uselargelanguagemodels,faggioli2023perspectives,macavaney2023one,thomas2023large}, they also highlight a lack of calibration between automated relevance judgments and human assessments, with LLM-based assessment often assigning higher grades on average~\cite{upadhyay2024umbrela}.


\paragraph{Alignment with Human labels. } Discrepancies exist between human and LLM-based relevance judgments. For instance, humans are found to be generally more strict in defining relevance compared to LLMs \cite{upadhyay2024largescalestudyrelevanceassessments,upadhyay2024umbrela,meng2024query}. These differences are also reflected in statistical agreement measures, such as Cohen's Kappa ($\kappa$).
When comparing human and LLM assessments, \citet{faggioli2023perspectives} report binary $\kappa$ values between 0.07 and 0.49 (slight to moderate). 
\citet{upadhyay2024umbrela} report  $\kappa$ agreement for binary and graded relevance judgment between LLM and human annotations. For binary, they report $\kappa$  between 0.4 and 0.5 (moderate) and for four-point values between 0.3 and 0.4.
\citet{LLMJudge} report an experiment in which seven independent research groups applied their prompts and LLM-based assessment tools to label a common set of query/document pairs on a four-point graded relevance scale.
No group achieved a $\kappa$ over 0.45, when compared with human labels, with all but one submission having $\kappa$ values below 0.4.
As a basis for comparison, human-human $\kappa$ values can be above 0.5 on binary assessment~\cite{10.1145/290941.291009}.

\paragraph{Agreement with System Rankings.}
In addition to a direct comparison of relevance labels, information retrieval researchers often evaluate relevance grading methods and evaluation metrics by computing Kendall's Tau ($\tau$) correlation between system rankings~\cite{voor98}.
Kendall's $\tau$ measures the rank correlation between two ordered lists, accounting for concordant and discordant pairs.
In the context of evaluation experiments, such as TREC tasks, Kendall's $\tau$ quantifies the consistency between the rank orderings induced by two different evaluation metrics across a set of submissions (i.e., ``runs'' or ``systems'').
In comparing official TREC binary relevance labels with an independently created set of labels on runs submitted to the TREC-6 adhoc retrieval task, \citet{voor98} report a Kendall's $\tau = 0.8956$, where the corresponding Kappa is 0.52~\cite{10.1145/290941.291009}.
Thus, if we are replacing human labels with LLM-generated labels, a Kendall's $\tau$ of $0.9$ could serve as a minimum benchmark for acceptability.

LLM-generated labels approach or exceed this standard when compared to human labels. \citet{faggioli2023perspectives} report $\tau = 0.86$ on TREC 2021 Deep Learning runs when comparing human labels with labels generated by GPT-3.5. Similarly, \citet{upadhyay2024umbrela} report $\tau$ values between $0.87$ and $0.94$ on TREC 2019 to 2023 Deep Learning runs when comparing human labels with labels generated by their \texttt{UMBRELA} assessment tool. Furthermore, \citet{upadhyay2024largescalestudyrelevanceassessments} report $\tau = 0.89$ on TREC 2024 runs using the same tool.

\paragraph{Alignment vs. Agreement.}
While previous studies have compared prompts for LLM-based graded assessment~\cite{LLMJudge}, to our knowledge, no prior work has systematically compared different LLM-based relevance assessment methods. Various research teams have reported results on different datasets, but due to the lack of comprehensive experiments, there has been no unified comparative analysis. This gap arises not only because relevant datasets and judgments are sometimes computationally/financially expensive to generate but also due to the absence of a clear and standardized methodology for making such comparisons. In this paper, we address these challenges by: (1) providing comprehensive LLM-based relevance judgments using five different methods and two different LLMs across four datasets, and (2) introducing two methodologies for comparing LLM-based relevance judgments—one based on alignment with human labels and the other on agreement with system rankings. By standardizing the evaluation process, we enable a more rigorous and reproducible comparison of different LLM-based assessment approaches.


\section{Relevance Assessment Methods}
\label{sec:methods}
In this section, we present a detailed explanation of the LLM-based relevance assessment methods compared in this paper.
To ensure replicability and to minimize potential bias introduced by prompt engineering, we have relied as much as possible on the tools and prompts developed by prior work. We note that all the experiments have been conducted with both GPT-4o and Llama3.2b with a temperature of 0. All the prompts, code and data for both models are available at our GitHub Repository.

\subsection{Binary Relevance Assessment}

Binary relevance assessment is a straightforward approach assessing relevance, which has been widely employed within the information retrieval community for much of its history~\cite{cleverdon1991significance}.
It forms the foundation for standard evaluation metrics like recall and average precision. 
A document \( d \) is evaluated to determine whether it 
satisfies the information need underlying a query \( q \). Given a query \( q \) and a document \( d \), a \texttt{Binary} relevance function \( \mathcal{B}(q, d) \) maps the query-document pair to a label \( r \):
\begin{small}
    \begin{equation}
    \mathcal{B}(q, d) = r, \quad \text{where } r \in \{0, 1\}
\end{equation}
\end{small}
Here, \( r = 1 \) indicates relevance, while \( r = 0 \) signifies non-relevance.

Binary judgments, while easy to understand and interpret 
\cite{manning2009introduction,kekalainen2005binary,kekalainen2002using}, can oversimplify the relevance spectrum, failing to capture more nuanced degrees of relevance. 
\citet{faggioli2023perspectives} were among the first to leverage LLMs for binary relevance judgments. For \texttt{binary} assessment we reproduce their prompt, which explicitly instructs the LLM to act as a TREC assessor when determining if a document is relevant to a query. Their prompt has previously shown high agreement with human judgments on datasets including REC-DL 2021 \cite{trecdl2021} and TREC-8 \cite{hawking1999overview}.

\subsection{Graded Relevance Assessment}

Early efforts in IR, highlighted the need for richer annotations to better evaluate retrieval systems~\cite{buckley2004retrieval,hawking1999overview,voorhees2000report}.
As such, graded relevance judgment extend binary relevance by assigning a relevance grade \( r \) from a predefined set of levels, typically ranging from 0 (non-relevant) to \( r_h \) (highest relevance).
For a given query \( q \) and document \( d \), a graded relevance function $\mathcal{G}(q, d)$ maps the query-document pair to an integer \( r \), where:
\begin{small}
\begin{equation}
\mathcal{G}(q, d) = r, \quad \text{where } r \in \{0, 1, \dots, r_h\}.
\end{equation}
\end{small}
The number of levels (\( h + 1 \)) is often 4 or 5 in standard test collections.
For instance, the TREC datasets use scales ranging from 0 (not relevant) to 3 (perfectly relevant) \cite{trecdl2019}.
Graded relevance judgments enables finer-grained evaluation metrics such as nDCG, with the goal of making evaluation better aligned with real-world user satisfaction, where not all relevant documents are equally useful.
However, graded relevance suffers from problems of subjectivity, where assessors may disagree on the label to assign to a particular document.
As a result, graded relevance requires more detailed guidelines and training for assessors.

As LLMs became prominent, several researchers explored their use in generating graded relevance judgments~\cite{arabzadeh2024adaptingstandardretrievalbenchmarks,alaofi2024generative,abbasiantaeb2024uselargelanguagemodels,meng2024query,thomas2023large}. Notably, Bing researchers announced that they had replaced human annotators with LLMs for some production-level relevance judgments, demonstrating the potential of LLMs to replace or augment human assessment~\cite{thomas2023large}.
The \textit{\texttt{UMBRELA}} method \cite{upadhyay2024largescalestudyrelevanceassessments}, provides open-sourced prompts replicating Bing's LLM-based relevance judgment system, demonstrating high agreement with system rankings, particularly in the TREC RAG 2024 evaluation. 
For this study, we adopted \texttt{UMBRELA}\footnote{\url{https://github.com/castorini/umbrela/tree/main}} prompts and implementation with a zero-shot setting.

\subsection{Nugget-based Relevance Assessment}

To facilitate the assessment process, some evaluation methods decompose the information need underlying a query.
Different papers give different names to the products of this decomposition, including ``facets'', ``subtopics'', ``rubrics'', and ``nuggets''. To maintain consistent terminology throughout this paper, we will refer to them as ``nuggets''. 
By considering relevance of the nuggets that a document satisfies, we can determine not only if a document is relevant to a query but also to what extent, and from which perspective, it fulfills it. 
Prior studies \cite{rajput2011nugget,pavlu2012ir,marton2006nuggeteer} have shown that nugget-based frameworks improve the interpretability of relevance judgments in information retrieval systems.
For this study, we adopt two distinct LLM-based relevance assessment approaches within the larger nugget-based framework:
1) A {\em Document-Agnostic} approach called \texttt{Exam} \cite{farzi2024pencils,sander2021exam} where nuggets are extracted solely based on the main query, and 2) a  {\em Document-Dependent}  method called \texttt{AutoNuggetizer} \cite{pradeep2024initial}, which extracts nuggets from documents which are known to be relevant to the query. 

\subsubsection{Document Agnostic (\texttt{Exam})}
Older TREC tracks employed subtopics manually extracted from queries to facilitate relevance judgment tasks~\cite{clarke2009overview,zhai2015beyond}.
Under the {\texttt{Exam}}\footnote{\url{https://github.com/laura-dietz/flan-t5-exam-appendix}} method \cite{farzi2024pencils,sander2021exam}, an LLM is prompted to generate nuggets based on the query alone~---~independent of any retrieved documents. This document agnostic approach avoids biases potentially introduced by using relevant documents as the source of nuggets but also limits nuggets to those that can be generated by the LLM without any additional context. 

\paragraph{\texttt{Exam} Nugget Generation:} For a given query \(q\), \texttt{Exam} derives 10 nuggets, denoted as $\mathcal{N}_\mathcal{E}(q) = \{ N^q_1, N^q_2, \ldots, N^q_{10} \} $.
The prompts used by \texttt{Exam} require the extracted nuggets to be concise and insightful questions, which together are intended to provide complete coverage for the query.

\paragraph{Nugget-Based Assessment} After the nuggets are generated, \texttt{Exam} determines the extent to which a given document satisfies each nugget.
Both binary and graded judgments may be employed at the per-nugget level.
\begin{itemize}[leftmargin=*]
    \item \texttt{Exam}-Binary: For document \(d\), an LLM is prompted to make a binary judgment with respect to each nugget \(N^q_i\), where \(B(N^q_i, d)\) is 1 if the document is relevant to \(N^q_i\), and 0 otherwise.
    These binary judgments are then aggregated to give a score for the document as follows:
    \begin{small}
    \begin{equation}
    \mathcal{E}_b(q, d) = \frac{1}{| \mathcal{N}_\mathcal{E}(q) |} \sum_{i=1}^{|\mathcal{N}_\mathcal{E}(q)|} \mathcal{B}(N^q_i, d)
    \end{equation}
    \end{small}
    \item \texttt{Exam}-Graded: As proposed by \citet{farzi2024pencils}, an LLM is prompted to assign an integer relevance grade \(r \in [1, 5]\) to each nugget \(N^q_i\) indicating the degree to which the document \(d\) satisfies it.
    We aggregate these per-nugget relevance grades in two ways, which are discussed separately below.
\end{itemize}

\paragraph{Graded Nugget Aggregation}
After \texttt{Exam}-Graded assigns a per-nugget relevance grade, these grades can be aggregated in two way to produce an overall score for a document:
\begin{itemize}[leftmargin=*]
    \item {\texttt{Exam}-Graded Max:} This is the original aggregation proposed by~\citet{farzi2024pencils} which selects the maximum score across all nuggets:

\begin{small}
\begin{equation}    
\mathcal{E}_{g_{\text{max}}}(q,d) = \max_{i=1}^{|\mathcal{N}_\mathcal{E}(q)|} \mathcal{E}_g(N^q_i, d) 
\end{equation}
\end{small}

\item {\texttt{Exam}-Graded Mean:} In addition, we explore a second aggregation function that computes an average score across nuggets:
\end{itemize}
\begin{small}    

\begin{equation}  
\mathcal{E}_{g_{\text{mean}}}(q,d) = \frac{1}{|\mathcal{N}_\mathcal{E}(q)|} \sum_{i=1}^{|\mathcal{N}_\mathcal{E}(q)|} \mathcal{E}_g(N^q_i, d)
\end{equation}
\end{small}


\subsubsection{Document Dependent (\texttt{AutoNuggetizer})}
This method generates nuggets based on both a query and relevant documents. Unlike the query-only approach in \texttt{Exam}, \texttt{AutoNuggetizer} \cite{pradeep2024initial} anchors nugget generation to documents that were judged as relevant. This coupling provides richer contextual grounding for the nuggets but may also introduce potential biases toward the source documents.
The overall approach is similar to that of \citet{10.1145/2124295.2124343}, who proposed a nugget-based test collection created through human assessment. We note that for implementing all of the following steps, we used prompts provided by \texttt{AutoNuggetizer}~\cite{pradeep2024initial}.

\textit{Nugget Generation.}
Given a query $ q$ , the nugget generation process in \texttt{AutoNuggetizer} considers a set of documents $ D_q$  that have been judged relevant to the query. We denote the set of nuggets generated by an LLM with \texttt{AutoNuggetizer} as
\[
\mathcal{N}_A(q, D_q) = \{N^q_1, N^q_2, \dots, N^q_n\},
\]
where $ n < 30$  and each document $ d \in D_q$  satisfies $  \mathcal{B}(d, q) > 0 $ or $\mathcal{G}(d,q)> 0$. This ensures that nuggets are derived exclusively from documents deemed relevant to the query $q$ .

The nugget generation process is an iterative process that analyzes all relevant documents \(d \in D_q\). This process is intended to ensure that the generated initial nuggets are neither too broad nor overly specific. Nuggets capturing similar concepts are recursively merged to maintain clarity and reduce redundancy. If the total number of nuggets exceed the limit (30) , the process involves merging redundant nuggets or removing less important ones. 

\textit{Nugget Importance.}
After nugget generation, each nugget is categorized based on their importance into one of two groups: (1) \textit{Vital} nuggets, which are essential for a good response, and (2) \textit{Okay} nuggets, which are useful but not critical for completeness. Further, all nuggets are sorted by their importance, and the top 20 most important nuggets are selected. Some queries may result in fewer than 20 nuggets, and in certain cases, nuggets may be truncated during this process.
The final nugget set for query \(q\) is denoted as:
\[
\mathcal{N}_{A,\mathrm{sorted}}(q, D_q) = \{(N^q_1, I_1), (N^q_2, I_2), \ldots, (N^q_n, I_n)\},
\]

where \(N^q_i\) represents a nugget, and \(I_i\) is its importance, either \textit{vital} or \textit{okay}. $N_i$ is as important as or more important than $N_j$ if $i < j.$

\textit{Nugget Assignment.}
Once the nugget list is finalized, the LLM assesses documents based on how well they support each nugget. Each document \(d\) and nugget $N_i$ is assigned $\mathcal{SP}(N^q_i,q)$ with one of the labels: \{\texttt{support}, \texttt{partial\_support}, \texttt{not\_support}\}.

\begin{table*}[t]
\centering
\caption{Scoring metrics for nugget evaluation. $n_v$ and $n_o$ stand for vital nuggets and okay nuggets, respectively. $s_i^v$ and $s_i^o$ indicate the scores on vital and okay nuggets, respectively. $n_a$, $n_v$, and $n_o$ represent the number of all nuggets, vital nuggets, and okay nuggets, respectively.}

\label{tab:scoring_metrics}
\scalebox{1}{ 
\begin{tabular}{llll}
\textbf{Metric} & \textbf{Definition} & \textbf{Metric} & \textbf{Definition} \\ \hline
All (\(A\)) = & 
$ \frac{\sum_i S_i}{n_\text{a}} $ &
All Strict (\(A_{\text{strict}}\)) = &
$ \frac{\sum_i SS_i}{n_\text{a}} $ \\
Vital (\(V\)) = &
$ \frac{\sum_i s_i^v}{n_v} $ &
Vital Strict (\(V_{\text{strict}}\)) = &
$ \frac{\sum_i ss_i^v}{n_v} $ \\
Weighted (\(W\)) = &
$ \frac{\sum_i s_i^v + 0.5 \cdot \sum_i s_i^o}{n_v + 0.5 \cdot n_o} $ &
Weighted Strict (\(W_{\text{strict}}\)) = &
$ \frac{\sum_i ss_i^v + 0.5 \cdot \sum_i ss_i^o}{n_v + 0.5 \cdot n_o} $ \\ \hline
\end{tabular}
}
\end{table*}

\textit{Aggregation.} The relevance of a document \(d\) for a query \(q\) is computed based on nugget assignments with two different scoring functions. $S_i$ is score of nugget $N^q_i$ for query $q$  and a more strict version of scoring $SS_i$ focuses only on nuggets that are either \texttt{support}ed or \texttt{not\_support}ed, omitting the \texttt{partial\_support}ed of as follows:

{\small
\begin{align}
S_i  &= 
\begin{cases} 
1,   & \mathcal{SP}(N^q_i, d) = \texttt{support} \\
0.5, & \mathcal{SP}(N^q_i, d) = \texttt{partial\_support} \\
0,   & \mathcal{SP}(N^q_i, d) = \texttt{not\_support}
\end{cases} \\
SS_i &= 
\begin{cases} 
1, & \mathcal{SP}(N^q_i, d) = \texttt{support} \\
0, & \mathcal{SP}(N^q_i, d) = \texttt{not\_support}
\end{cases}
\end{align}

\citet{upadhyay2024largescalestudyrelevanceassessments} propose six different aggregation functions, varying the focus between vital nuggets only or all nuggets, and distinguishing between stricter scoring (\(SS\)) and more flexible scoring (\(S\)). These aggregation functions are summarized in Table \ref{tab:scoring_metrics}, where three core scoring functions are defined, and by replacing \(S_i\) with \(SS_i\), the total expands to six scoring functions: In the paper we refer to them as \texttt{Nuggets All (A)} and \texttt{Nuggets All (A) Strict}, which assign weights to all nuggets; \texttt{Nuggets Vital (V)} and \texttt{Nuggets Vital (V) Strict}, which focus exclusively on vital nuggets and disregard the okay nuggets; and a weighted summation of vital and okay nuggets, \texttt{Nuggets Weighted (W)} and \texttt{Nuggets Weighted (W) Strict}, where greater weight is assigned to vital nuggets compared to less important ones.

\subsection{Pairwise Relevance Judgments}
Pairwise preferences judgments provide a simple way to capture human feedback that ensures better inter-assessor agreement~\cite{kyct13,cvs21,cbcd08,sz20,xie20,seifikar2023preference,clarke2023preference}. Human preference labels are also central to the reinforcement learning from human feedback (RLHF) process used to tune LLMs and other models~\cite{christiano2023deep}. Pairwise judgments are less sensitive to subjective biases and focus on ordinal relationships, which are often more consistent across assessors~\cite{10.1145/3451161,Yan_2022}. Pairwise relevance judgment evaluates the relative relevance of two documents with respect to a given query. Instead of assigning absolute relevance scores or grades, this method asks the question ``Which document is more relevant to the query?'' 

Formally, given a query $q$ and two documents \(d_a\) and \(d_b\), a pairwise function $\mathcal{P}(q, d_a, d_b)$ determines the relative preference between the documents. To avoid positional biases, for LLM-generated preference judgments we repeat the assessment by swapping $d_a$ and $d_b$. In case both $\mathcal{P}(q, d_a, d_b)$ and $\mathcal{P}(q, d_b, d_a)$ agreed on the output, we consider the output document to be more relevant than the other one. Otherwise, we consider outcome to be a \textit{tie} i.e., both document enjoy the same level of relevance.

Since there are no established LLM-based relevance assessment tools for pairwise preference judgment, we developed our own prompt based on the instructions given to human preference judges by \citet{10.1145/3451161}. The exact prompt and assessment script is included in our software release.
While complete judgments for \(n\) documents would require \(O(n^2)\) preference judgments, we follow the sampling process of \citet{10.1145/3451161} which reduces the total number of judgments to \(O(n)\), or more specifically to less than $P \cdot n$ judgments, where $P = 7$, the default in their code release.
For aggregation, we assign an overall score to a document by counting the number of pairings it wins, ignoring ties, similar to a Borda score.

\section{Methodology}
In this section, we detail methodologies for determining the degree to which LLM-generated relevance judgments reflect preferences imputed by human labels (Section \ref{sec:HA-explain}) and quantifying their agreement with system rankings (Section \ref{sec:SR-explain}).

\subsection{Alignment with Human Labels}
\label{sec:HA-explain}
For each of the methods described in Section~\ref{sec:methods}, we measure alignment with human labels by asking the following question:
\begin{quote}
\textit{When human assessment determines that one response to a query is better than another, how often does LLM-based assessment agree?}
\end{quote}
This approach allows us to compare methods that are based on different approaches to relevance judgment. 
Since some test collections use different relevance grades, and since some queries do not have documents at each relevance level, for each query we group the relevance judgments 
into three categories as follows, simplifying and unifying our analysis:
\begin{itemize}[leftmargin=*]
\item \textit{Best known}:
For each query, these are the relevant judged documents at the highest grade of relevance for that query. Given a 4 scale level of relevance (0-3) Some queries may have no relevant document with grade 3, or even grade 2, although we assume all queries have at least one judged non-relevant document.
\item \textit{Acceptable}:
For each query, these are the relevant documents that fall between the grade of the \textit{best known} answer and the judged non-relevant ones (grade~0).
\item \textit{UnAcceptable:}
For each query, these are the documents judged as not relevant to the query (grade~0).
\end{itemize}
Each method is measured in terms of its ability to make assessments that are consistent with the order imputed by these categories.
If an LLM-based relevance assessment method orders documents consistently with these categories, it is said to ``Agree'' with human labels.
If an LLM-based relevance assessment method orders documents inconsistently with these categories, it is said to ``Disagree'' with human labels.
If an LLM-based relevance assessment method assigns the same score to documents from different categories, it is considered a ``Tie''.

We make comparisons between documents in the three categories as follows:
1) Best vs.\ Acceptable, 
2) Acceptable vs.\ UnAcceptable, and
3) Best vs.\ UnAcceptable.
We do not make comparisons between document in the same category (e.g., Best vs.\ Best).
Even if an LLM-based method would indicate a difference between two documents in the same category, we do not know if this difference is due to the LLM recognizing a finer distinction in relevance or due an error on the part of the LLM.
Given the limited number of relevance labels, we do not expect that documents assigned the same label are truly ``tied'' in relevance, and we do not demand that an LLM-based relevance assessment method assign them the same score.
In measuring alignment, we focus on comparison where the human labels indicate a clear difference in relevance.
Like Kendall's $\tau$, this approach to measuring alignment with human labels focuses on the ordering of paired items (concordant vs.\ discordant) but with ties handled in an application specific way.

\subsection{Agreement with System Rankings}
\label{sec:SR-explain}
In addition to ensuring relevance judgments are aligned  with human labels,  relevance judgments must also enable effective system ranking. Different relevance judgment methods~---~such as binary, graded, and nugget-based~---~produce outputs on varying scales depending on their aggregation strategies. For instance, \texttt{Exam} can yield different ranges of outputs depending on the aggregation strategy used. \texttt{AutoNuggetizer}  provides six distinct aggregation methods, each with its own range of values and scale.

To allow a direct comparison between relevance assessment methods, despite their varying scales, we adopt the \textit{compatibility} evaluation metric introduced in \citet{clarke2020offline}. Compatibility offers a robust framework for comparisons between relevance judgment methods by focusing solely on the relative ordering of query-document pairs rather than their absolute scores.
Compatibility is agnostic to the numerical scale used to express relevance assessments, enabling comparisons across diverse methods. Further, it allows direct comparisons between different relevance judgment methods without requiring normalization or recalibration of scores.

Let us consider an ideal ranking \textit{set} $I$, which represents the preferred ordering of documents for a query. It is constructed from a set of \textit{effectiveness levels} $\{L_1, L_2, \ldots, L_T\}$, where $L_T \succ L_{T-1} \succ \ldots \succ L_1$
and $L_T$ contains the most relevant documents. Documents within the same level \( L_i \) are considered equivalent and can appear in any order. Non-relevant documents belong to \( L_0 \) are excluded from the ideal ranking. As such, we can obtain an ideal set of ranking \( I_\mu \) for any given relevance judgment method \( \mu \) by sorting the documents of a given query based on their different levels of relevance. The number of levels in \( I_\mu \) may vary across methods, as  \( \mu \) might produce different ranges of relevance.

Given an actual ranking \( R \), the ideal ranking set \( I_\mu \) represents the set of ranking permutations created by ordering documents within each relevance level \( L_i \) according to their positions in \( R \).
By aligning the document order within each level to various permutations consistent with \( R \), \( I_\mu \) ensures the ideal ranking is both closely aligned with \( R \) and robust to ties within \( R \). \( I' \) is a permutation in \( I_\mu \) where the relevance levels are correctly ordered, but documents within the same level may appear differently in different permutations.

We can measure the similarity between $R$ and any $I' \in I$ with any appropriate rank similarity measure. Following~\citet{10.1145/3340531.3411915}, we use Rank Biased Overlap (RBO) \cite{webber2010similarity}, a rank similarity measure that emphasizes agreement at higher ranks. RBO is flexible in controlling the weight assigned to deeper ranks, with larger \( p \) values placing greater emphasis on lower-ranked documents. As suggested in \cite{clarke2021assessing}, we set \( p = 0.9 \). Finally the compatibility score is defined as the maximum RBO value between \( R \) and any ranking permutation $I'$ in the set of ideal rankings \( \mathcal{I}_\mu \):
\begin{equation}
\text{Compatibility}(R, I_\mu) = \max_{I' \in \mathcal{I}_\mu} \text{RBO}(R, I')
\end{equation}
In summary, compatibility measures the similarity of a ranking to the most similar ideal permutation of relevance judgments.




\begin{table}[t!]
    \centering
    \caption{Statistics of datasets used in our experiments.}
    \label{tab:dataset_stats}
    \scalebox{0.9}{
    \begin{tabular}{lp{3em}p{4em}p{3em}p{4em}p{5em}}
        \hline
        \textbf{Dataset}  & \textbf{Queries} & \textbf{Non-relevant} & \textbf{Related} & \textbf{Relevant} & \textbf{Perfectly Relevant} \\
        \hline
         DL 2019  & 43  & 5,158 & 1,601 & 1,804 & 597 \\
         DL 2020  & 45  & 7,780 & 1,940 & 1,020 & 646 \\
         DL 2021 & 57  &  4,338 & 3,063 &  2,341&  1,086\\
        ANTIQUE    & 200 & 1,642 & 2,417 & 1,196 & 1,334 \\
        \hline
    \end{tabular}}
\end{table}

\section{Experiments and Results}
This section provides the details of our experiments.
While the results presented below were obtained using GPT-4o, results for Llama3.2 are also available in our repository.

 \subsection{Datasets}

We conduct experiments on three passage ranking datasets from the TREC Deep Learning (DL) Tracks i.e.,  {TREC DL 2019} \cite{trecdl2019}, {TREC DL 2020} \cite{trecdl2020}, and {TREC DL 2021} \cite{trecdl2021}, along with the {ANTIQUE} dataset \cite{hashemi2020antique}, a benchmark for non-factoid question answering.
DL-19 and DL-20 utilize the MS MARCO v1 collection \cite{nguyen2016ms}, containing more than 8.8 million passages; DL 2021 utilizes the updated MS MARCO {v2} collection with approximately 138 million passages. 
For all TREC DL datasets, human annotators from the National Institute of Standards and Technology (NIST) conducted relevance assessments using a four-point graded scale including Perfectly relevant (3), Highly relevant (2), Related (1) and Irrelevant(0). 
Relevance judgments were collected through a standard pooling strategy, where the top documents from submitted runs were pooled for human assessment.

\begin{figure*}[t!]
    \centering
    \includegraphics[width=0.9\linewidth]{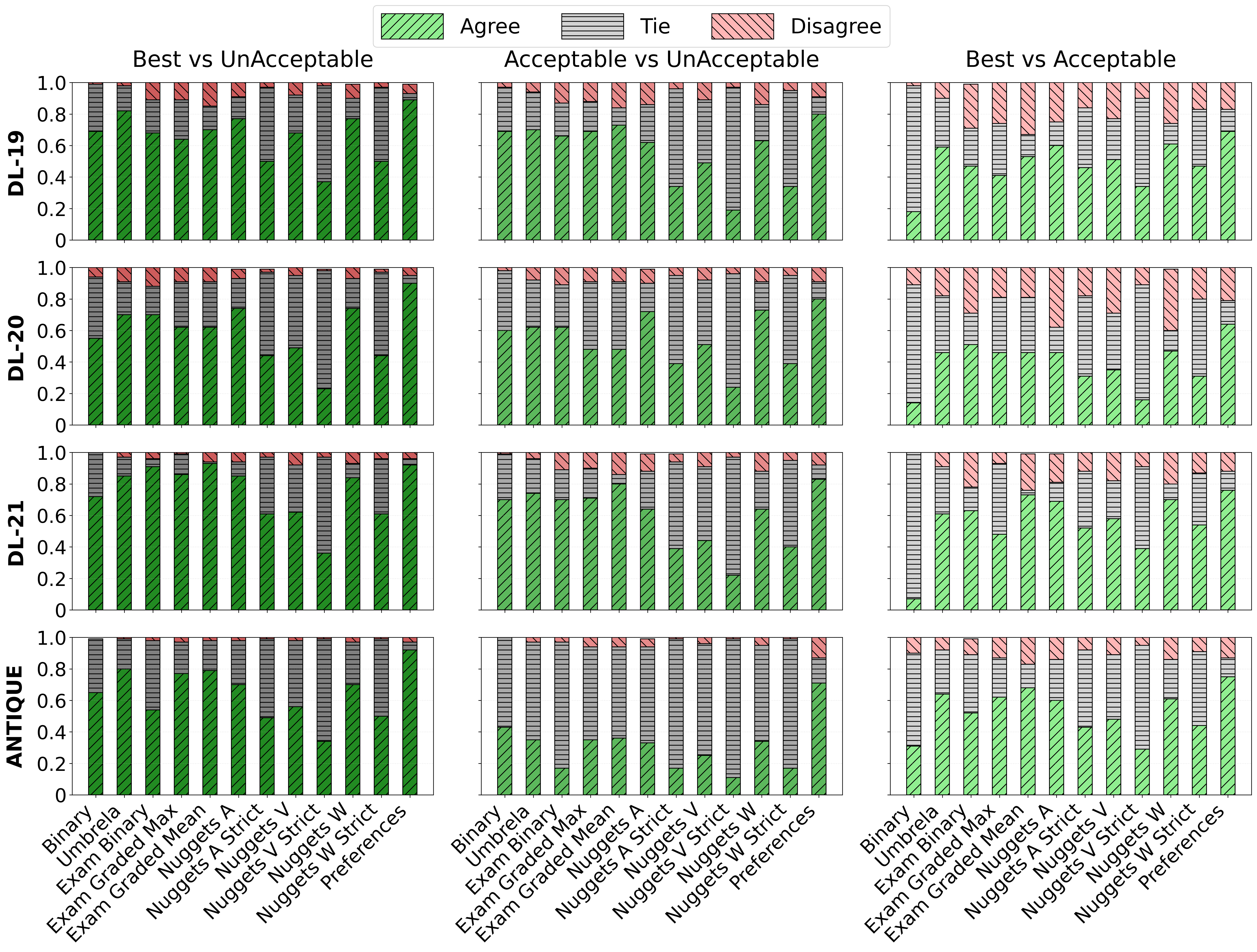}
\caption{Human alignment results for different LLM-based Relevance Judgment methods on DL-19, DL-20, DL-21, and ANTIQUE (top to bottom rows). Within each dataset, comparisons are shown across different relevance categories: Best vs. UnAcceptable, Acceptable vs. UnAcceptable, and Best vs. Acceptable (leftmost to rightmost columns). Darker colors reflect greater ease in distinguishing between the two categories of relevance.}
    \label{fig:human_alignment}
\end{figure*}

We extend our experiments beyond just TREC-style relevance judgment assessment by conducting experiments on a non-factoid question answering benchmark, ANTIQUE dataset \cite{hashemi2020antique}. ANTIQUE consists of {2,626} non-factoid questions sourced from Yahoo! Answers, accompanied by {34,011} manually annotated relevance judgments. The test set includes {200} queries, judged by multiple annotators using a four-level relevance scale.
To ensure annotation quality, each answer in ANTIQUE dataset was evaluated by three annotators, with disagreements resolved through additional annotation rounds or expert adjudication. 

Table~\ref{tab:dataset_stats} summarizes the datasets used in our experiments, detailing the number of queries and the distribution of relevance judgments across different graded levels.
Evaluating both the TREC Deep Learning datasets and the ANTIQUE dataset allows us to assess the generalizability of our approach across diverse retrieval tasks.

\subsection{Alignment with Human Labels}

\label{results:ha}

In this section, we present the results of the experiments described in Section \ref{sec:HA-explain}, where we measure the average alignment of each relevance judgment method with human labels across various levels of relevance. The results are based on the methods described in Section \ref{sec:methods} and are evaluated on four datasets. Each bar in Figure~\ref{fig:human_alignment} is stacked
to show the percentage of pairs for which the method agrees with human judgment (green), results in a tie (gray), or disagrees (red). The ``Agree'' category indicates that the method ranks a document from a higher relevance category above a document from a lower relevance category, matching the order imputed by human labels. The ``Tie'' category represents cases where the method assigns the same score to both documents -- even though they were from different levels of relevance as annotated by human -- indicating uncertainty or indistinguishability for the relevance judgment method. Finally, the ``Disagree'' category represents cases where the method assigns a higher score to a document from a lower relevance category. Higher agreement percentages indicate better alignment with human labels.

The columns in figure \ref{fig:human_alignment}, moving from left to right, correspond to comparisons with decreasing relevance distinctions (e.g., Best vs. UnAcceptable, to Best vs. Acceptable). Methods generally achieve higher agreement rates when the relevance distinction between categories is more pronounced (the  leftmost column). As expected, alignment is significantly higher for Best vs. UnAcceptable comparisons  compared to the other two columns. 
As we move to finer distinctions, such as Best vs. Acceptable, the agreement rate decreases, while ties and disagreements increase.

\begin{figure*}[]
\centering
  \vspace{-1em}
  \includegraphics[clip, trim=0cm 5.8cm 6cm 0cm,scale=0.6]{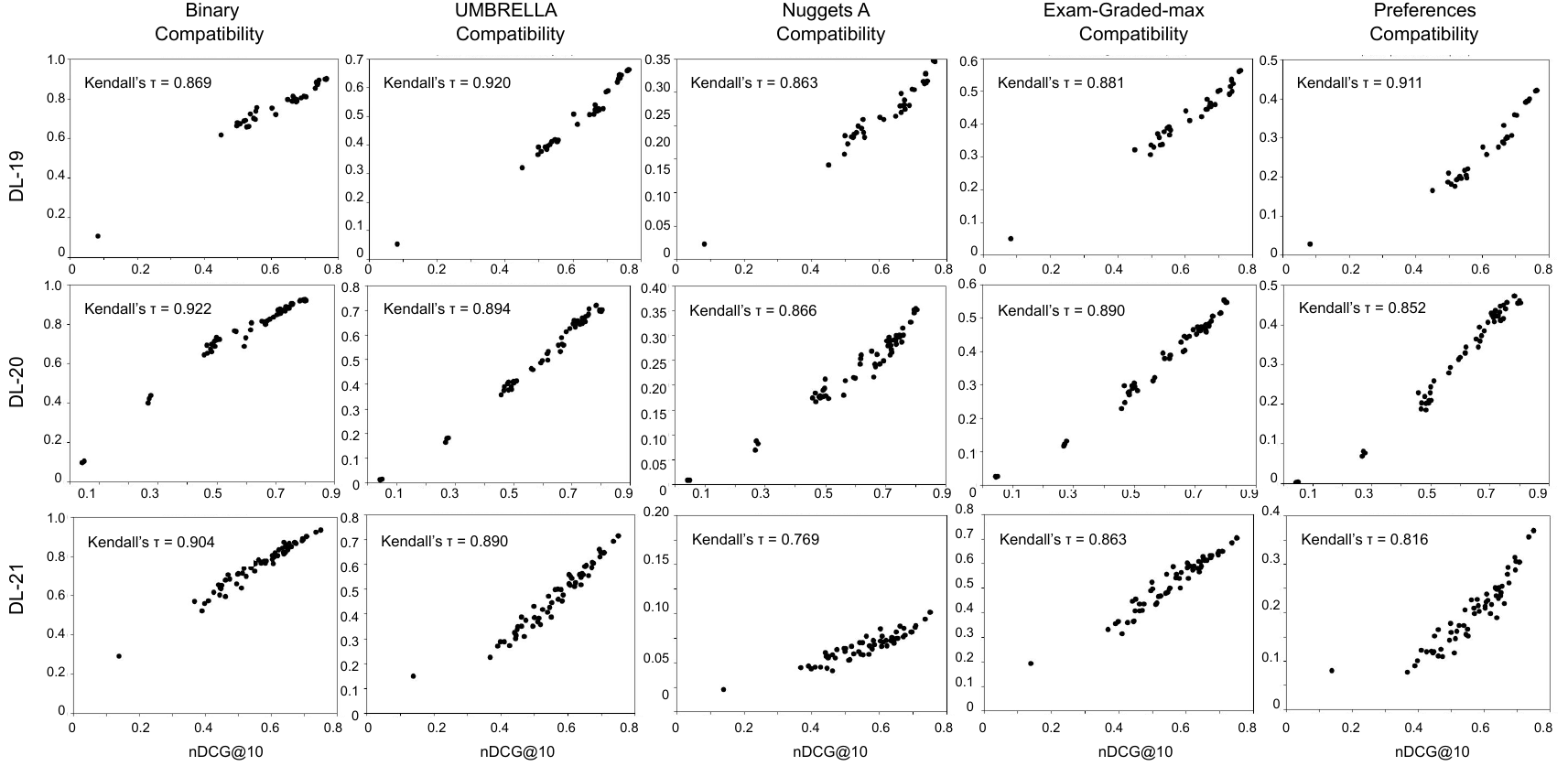}
\caption{Compatibility (LLM assessment) vs nDCG@10 (human assessment) for the relevance assessment methods with the highest Kendall correlation (see Table \ref{tab:metrics}) on runs submitted to TREC DL-20 and DL-21.
Plots for all assessment methods and datasets are included in the GitHub repo.}
\vspace{-1em}
\label{fig:compatibility}
\end{figure*}

The \texttt{preferences} method consistently achieves the highest agreement rates across all datasets, indicating strong alignment with human labels. For example, in the Best vs. UnAcceptable comparison, \texttt{preferences} achieves an average agreement of 91\% across all datasets. Even in more challenging scenarios e.g., as Best vs. Acceptable, \texttt{preferences} still maintains an average agreement of 71\%, outperforming other methods. \texttt{UMBRELA} and \texttt{Nuggets All} are the next best-performing methods, with average agreement rates of 79\% and 77\% for Best vs. UnAcceptable across all four datasets.

\begin{sloppypar}
Nugget-based methods such as \texttt{Nuggets A} and \texttt{Nuggets W} exhibit acceptable alignment on the TREC DL datasets but struggle on the ANTIQUE dataset. This performance drop on ANTIQUE may be attributed to its non-factoid nature, where broader and more diverse nuggets are possible. In fact, nugget-based methods display high variability, often excelling on one dataset while underperforming on another, suggesting a lack of consistency. 
\end{sloppypar}
The ANTIQUE dataset also exhibits a higher proportion of ties across most methods. This may be due to the non-factoid nature of the dataset, where answers can be valid from multiple perspectives.
\begin{table}[t]
\caption{Kendall $\tau$ correlations for different relevance judgments with against compatibility (p = 0.9). As a basis for comparison, the ``original human'' values show the extent to which compatibility correlates with nDCG@10 under the original human labels.
Highest correlation is bolded;
second-highest is underlined.}
    \centering
    \begin{tabular}{lccc}
        \toprule
        Metric & DL-19 & DL-20 & DL-21 \\
        \toprule
        Original human & 0.953 & 0.956 & 0.916 \\
        \midrule
        \texttt{Binary} & 0.869 & {\bf 0.922} & {\bf 0.904} \\
        \texttt{UMBRELA }& {\bf 0.920} & \underline{0.894} & \underline{0.890} \\
        \texttt{Exam-Binary} & 0.794 & 0.881 & 0.711 \\
        \texttt{Exam-Graded$_{\texttt{max}}$} & 0.881 & 0.890 & 0.863 \\
        \texttt{Exam Graded$_{\texttt{mean}}$} & 0.863 & 0.890 & 0.832 \\
        \texttt{Nuggets A} & 0.863 & 0.866 & 0.769 \\
        \texttt{Nuggets A strict} & 0.857 & 0.791 & 0.810 \\
        \texttt{Nuggets V} & 0.839 & 0.747 & 0.712 \\
        \texttt{Nuggets V strict}& 0.836 & 0.685 & 0.712 \\
        \texttt{Nuggets W }& 0.860 & 0.829 & 0.760 \\
        \texttt{Nuggets W strict} & 0.824 & 0.767 & 0.794 \\
        \texttt{Preferences} & \underline{0.911} & 0.852 & 0.816 \\
        \bottomrule
    \end{tabular}
        \vspace{-1em}
    \label{tab:metrics}

\end{table}

While \texttt{Binary} relevance judgments are simple and interpretable, they fail to capture fine-grained distinctions, resulting in high percentage of ties and lower alignment in challenging comparisons.  \texttt{Binary} relevance, especially in Best vs. Acceptable comparisons, shows the highest average tie rate (77\%), likely because both answers meet a broad relevance threshold. 
In contrast, graded and nugget-based methods provide more nuanced assessments, but their performance varies significantly depending on the dataset and category distinction. These observations underscore the importance of selecting the appropriate relevance judgment method based on the dataset and task requirements. Methods such as \texttt{preferences} demonstrate superior consistency and alignment with human judgment across datasets, making them a robust choice for automated relevance evaluation. 

\subsection{Agreement with System Rankings}

We analyze the agreement of different relevance judgment methods using the compatibility metric introduced in Section \ref{sec:SR-explain}. This analysis is conducted with respect to the official evaluation metric of the TREC DL datasets, nDCG@10, calculated based on the original human-annotated relevance judgments (original qrels). 
We treat the system ranking under nDCG@10 as the ``gold standard'' comparing the system rankings under other methods to its ranking.

Table~\ref{tab:metrics} presents Kendall $\tau$ correlations between nDCG@10 and the compatibility metric across all submitted runs for the TREC Deep Learning Passage Ranking tasks. The runs were obtained from the \url{https://trec.nist.gov/} website. For this section, we conduct experiments exclusively on the TREC DL datasets, as they provide sufficient runs for comparing different systems and performing system rankings. In contrast, the ANTIQUE dataset had limited available runs, and we could not study system ranking on it.

\begin{figure*}[t!]
    \centering
    \includegraphics[width=\linewidth]{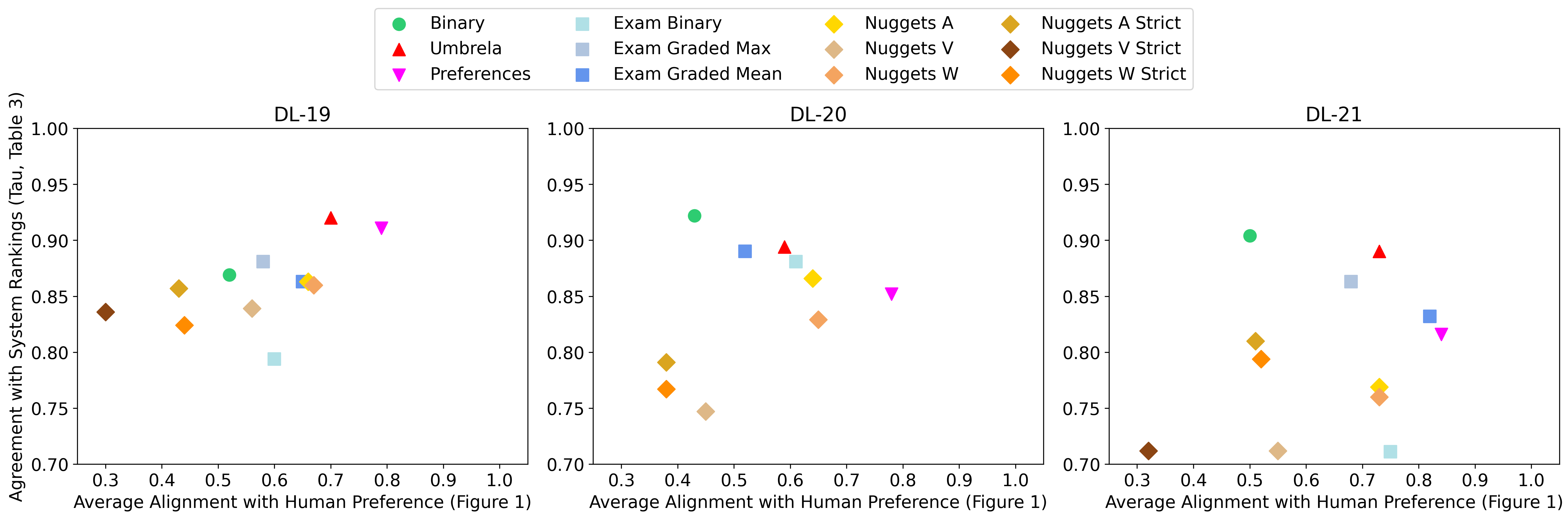}
    \caption{Comparing average alignment with human preferences from Figure \ref{fig:human_alignment} with system ranking agreement from Table \ref{tab:metrics} on different relevance judgment methods across three TREC DL datasets.}
    \vspace{-1em}
    \label{fig:alignment_vs_ranking}
\end{figure*}

To help validate the use of compatibility, the first line of the table compares compatibility measured with human labels to nDCG@10 measured with human labels.
This line indicates the extent to which compatibility can replace nDCG@10 for relevance assessment methods for which nDCG is not suited.
The original human labels demonstrated consistently high compatibility, with Kendall’s tau values exceeding 0.9 across all three datasets (DL-19, DL-20, and DL-21). This level of correlation validates compatibility for comparing relevance judgment methods.

Among all methods, pointwise approaches such as \texttt{Binary} and \texttt{UMBRELA} achieve the highest agreement with human-based system rankings. \texttt{Binary} relevance judgments exhibited the strongest correlations on DL-20 and DL-21, while \texttt{UMBRELA} was the leader for DL-19. These results emphasize the reliability of traditional binary and graded methods in capturing system-level agreement, even when the assessment is performed by an LLM, but may also reflect the fact that the gold standard is also based on traditional relevance grades.
Despite the inherent challenges in comparing pairwise \texttt{preferences} with system rankings, due to their lack of absolute values for traditional metrics like nDCG, our systematic evaluation framework allowed these judgments to be integrated and compared effectively. Remarkably, pairwise \texttt{preferences} showed high correlations, emerging as the runner-up on DL-19 and maintaining competitive performance on the other two datasets.

Nugget-based methods exhibited variability across datasets. Document-agnostic approaches, such as \texttt{Exam-Binary}, tended to have higher correlations on DL-20, while document-dependent approaches showed slightly better alignment on DL-19. However, their performance was generally less consistent than that of pointwise and pairwise methods.
As a further illustration, we present scatter plots for \texttt{Binary}, \texttt{UMBRELA}, \texttt{Exam-Graded$_{\texttt{max}}$}, \texttt{Nuggets A} and Pairwise \texttt{Preferences}, identified as the top three best-performing methods in Table \ref{tab:scoring_metrics}. Figure \ref{fig:compatibility} presents scatter plots of all submitted runs from DL-20 and DL-21 for these three methods. Due to space constraints, we only showcase these scatter plots; however, scatter plots for the remaining methods and datasets are available on our GitHub repository. 
The results underline the effectiveness of compatibility as a robust metric for comparing diverse relevance judgment strategies, enabling meaningful comparisons across methods with varying output scales.

\subsection{Alignment vs.\ Agreement}

In this section, we integrate the findings from Sections 5.2 and 5.3 to examine how human alignment and system ranking agreement interact and correlate. 
Figure \ref{fig:alignment_vs_ranking} illustrates this relationship by plotting agreement with system ranking (y-axis) vs.\ average alignment with human labels (y-axis).
Average alignment is computed by (macro-)averaging the ``Agree'' values across the three relevance comparisons in Figure~\ref{fig:human_alignment}~---~Best vs.\ UnAcceptable, Acceptable vs.\ UnAcceptable, and Best vs.\ Acceptable.
The agreement with system rankings is measured by the Kendall’s Tau correlation between compatibility and nDCG@10, as reported in Table~\ref{tab:metrics}. 


From this figure, we observe that Pairwise \texttt{Preferences} consistently achieve the highest human alignment across datasets but shows comparatively lower agreement with system rankings. On the other hand, \texttt{Binary} relevance judgments and \texttt{UMBRELA} methods demonstrate strong system ranking agreement but do not always achieve equally high alignment with human labels. These results suggest that methods prioritizing alignment with human labels may not inherently optimize for agreement with system rankings, and vice versa.

\begin{sloppypar}
    We also note that the aggregation strategies in nugget-based approaches play a significant role. \texttt{AutoNuggetizer (Nuggets)} and \texttt{Exam}-based approaches show varying degrees of alignment and agreement depending on the dataset and variations. For example, among the six variations of \texttt{nugget}-based relevance judgments, we observe substantial differences in their alignment with human preferences and system rankings. This variability may be driven by the choice of aggregation strategy and the weighting of nuggets based on their importance. 
\end{sloppypar}
Overall, these findings highlight the need to carefully select relevance judgment methods based on the specific objectives of the evaluation. If the goal is to maximize alignment with human labels, Pairwise \texttt{Preferences} may be the optimal choice. However, if agreement with system rankings is more critical, methods such as \texttt{UMBRELA} or \texttt{Binary} relevance judgments may be more suitable. 

\section{Concluding Remarks}

In this paper, we compare methods for LLM-based relevance assessment.
We base our comparison on two factors: 1) alignment with human labels and 2) agreement with system rankings.
For alignment with human labels, we adopt a novel approach of comparing pairwise agreement between LLM-generated assessment and human labels.
This approach focuses on determining whether LLMs correctly preserve the relative order of relevance between documents as inferred from human labels, even when the absolute scores may differ.
This approach enables fine-grained evaluation of how closely LLM judgments align with human labels, independent of scale or scoring differences.
For agreement with system rankings, we employ Kendall's $\tau$, a standard metric for measuring rank correlation.
To ensure a fair comparison across all relevance assessment methods \cite{bigdeli2023biasing}, we introduce compatibility as a unified evaluation scoring function. This function facilitates consistent comparisons across different relevance judgment methods, placing them on a level playing field.

Perhaps because they directly compare two documents, preference labels consistently provide the best alignment with human labels.
Perhaps because they mimic the graded assessment process, UMBRELA relevance labels consistently provide the best or second-best agreement with system rankings verses human labels.
To operationalize these observations, we might imagine a human auditing or confirming those UMBRELA labels that are inconsistent under LLM-generated preference labels.
We leave the exploration of this idea to future work.
Also, we hope to compare LLM-generated preferences judgments with human preference judgments.
While, we study specific relevance assessment approaches and test collection in this paper, our overall methodology can be applied to any assessment approach or test collection.
We report results using an open-source (Llama) model in our repo, and we plan to extend our study to additional assessment approaches in the future.

\bibliographystyle{ACM-Reference-Format}
\balance
\bibliography{sample-base}
\balance
\end{document}